\documentclass[aps,prl,twocolumn,superscriptaddress]{revtex4}
\usepackage{mathrsfs,amssymb,graphics,subfigure,threeparttable}
\usepackage{grap hicx}
\usepackage{amssymb}
\usepackage{enumerate}
\usepackage{amsmath}
\usepackage{fancyhdr}
\usepackage{bm}
\usepackage[squaren]{SIunits}
\usepackage{hyperref}
\usepackage{ulem}
\usepackage{textcomp}
\usepackage{color}
\usepackage{float}
\usepackage[authormarkup=none]{changes}

\begin{document}

\preprint{APS/123-QED}

\title{Positron beam loading and acceleration in the blowout regime of plasma wakefield accelerator}

\author{Shiyu Zhou}
\affiliation{Department of Engineering Physics, Tsinghua University, Beijing 100084, China}
\author{Weiming An}
\affiliation{Institute for Frontiers in Astronomy and Astrophysics, Beijing Normal University, Beijing 102206, China}
\affiliation{Department of Astronomy, Beijing Normal University, Beijing 100875, China}
\author{Siqin Ding}
\affiliation{Department of Engineering Physics, Tsinghua University, Beijing 100084, China}
\author{Jianfei Hua}
\affiliation{Department of Engineering Physics, Tsinghua University, Beijing 100084, China}
\author{Warren B. Mori}
\affiliation{University of Los Angeles, Los Angeles, California 90095, USA}
\author{Chan Joshi}
\affiliation{University of Los Angeles, Los Angeles, California 90095, USA}
\author{Wei Lu}
\email[]{weilu@tsinghua.edu.cn}
\affiliation{Department of Engineering Physics, Tsinghua University, Beijing 100084, China}

\date{\today}

\begin{abstract}
  Plasma wakefield acceleration in the nonlinear blowout regime has been shown to provide high acceleration gradients and high energy transfer efficiency while maintaining great beam quality for electron acceleration. In contrast, research on positron acceleration in this regime is still in a preliminary stage. We find that an on-axis electron filament can be self-consistently formed and maintained by loading an intense positron beam at the back of the electron beam driven blowout cavity. Via an analytic model and fully nonlinear simulations, we show this coaxial electron filament not only can focus the positron beam but changes the loaded longitudinal wakefield in a distinctly different way from electron beam loading in the blowout regime. Using simulations, we demonstrate that a high charge positron beam can be accelerated with tens of percent energy transfer from wake to positrons, percent level induced energy spread and several mm$\cdot$mrad normalized emittance, while significantly depleting the energy of the electron drive beam. This concept can be extended to simultaneous acceleration of electron and positron beams and high transformer ratio positron acceleration as well.
\end{abstract}

\maketitle


The demand for particle beams at energy frontier has motivated intense research on novel accelerator techniques with higher accelerating gradient and energy transfer efficiency. This is particularly so for the future electron-positron colliders because leptons have an ever-increasing synchrotron energy loss as the beam energy increases\cite{Lou2019,Assmann2020, Joshi2020}. Over the last two decades, the charged particle-driven Plasma wakefield accelerator (PWFA) has made series of breakthroughs in electron beam acceleration, that provides sustained acceleration gradients of 1-50GeV/m over meter scale\cite{Chen1985,Blumenfeld2007}, energy extraction efficiency tens of percent\cite{Litos2014} and energy spreads of several percent\cite{Wu2019,Pompili2021,Lindstrom2018Emit}. These groundbreaking results were obtained in the so-called blowout regime by using an ultrashort and high-peak-current electron beam to drive electrons out of a roughly spherical region and leave behind a copropagating ion cavity that focuses and accelerates a trailing electron beam\cite{Lotov2004,Lu2006POP,Lu2006PRL}. However, no equivalent results have been obtained for positron acceleration in this regime mainly because the volume where the wakefield is both accelerating and focusing for a positron beam is extremely small and exists only when the transversely blown-out plasma electrons are attracted by the ions to form an electron density spike\cite{Lotov2007,Wang2008,Wang2009}. Alternative solutions for positron acceleration utilize specific plasma structures such as a finite-width plasma\cite{Diederichs2019,Diederichs2020}, hollow plasma channel\cite{Silva2021,Zhou2021,Zhou2022} or employ a drive beam that has a complicated profile such as a doughnut shape laser or electron beam\cite{Vieira2014,Jain2015}, or a positron drive beam\cite{Corde2015a}. These tailored plasma structures and/or beam profiles pose extra demands on experiments and may induce additional instabilities. Furthermore, there is still no theoretical model for the nonlinear positron beam loading effect and even a heuristic model is important for developing intuition, scaling laws and beam quality optimization.

In this Letter, we show that loading a positron beam just behind the first bubble in the blowout regime can self-consistently induce an extended focusing region and a flattened  accelerating field. As in other positron acceleration schemes\cite{Diederichs2019,Zhou2021} the focusing field is caused by a narrow electron filament forming on the axis. We present the first theoretical framework for positron beam loading that reveals the electron filament also enables flattening of the wake. Using three-dimensional (3D) quasi-static particle-in-cell (PIC) simulations we demonstrate that positron beam loading in the nonlinear blowout regime can lead to high-efficiency and high-quality positron acceleration. 
In an unloaded blowout wake, sheath electrons between the end of the bubble (first bucket) and the front of the second bucket are transitioning from moving towards the axis to moving away. By placing an intense positron beam in this region, its space charge force attracts some sheath electrons back towards the axis forming an extended on-axis electron filament that overlaps with the positron beam.
This filament not only provides the necessary focusing force, but makes positron beam loading of the accelerating field distinct from electron beam loading in the blowout regime. We extend the approach of Lu et al.\cite{Lu2006POP,Lu2006PRL} to develop an analytical model to include the interplay between the electron filament and the bubble profile. This model reveals the possibility for a nonlinear wake driven by an electron beam to provide a flattened accelerating field and nearly linear focusing fields. Simulations show that it is able to simultaneously achieve $\sim$1\% induced energy spread, emittance preservation of several mm$\cdot$mrad and energy transfer efficiency over 20\% from wake to the positron beam. Furthermore, an additional electron bunch can be loaded in the same wake in the first bubble which can lead to the improvement of the positron beam quality and overall beam loading efficiency. The possibilities of positron acceleration with a high transformer ratio (HTR) and even simultaneous HTR electron and positron acceleration with a single driver are illustrated.

\begin{figure}[ht]
  \includegraphics[width=8.6cm]{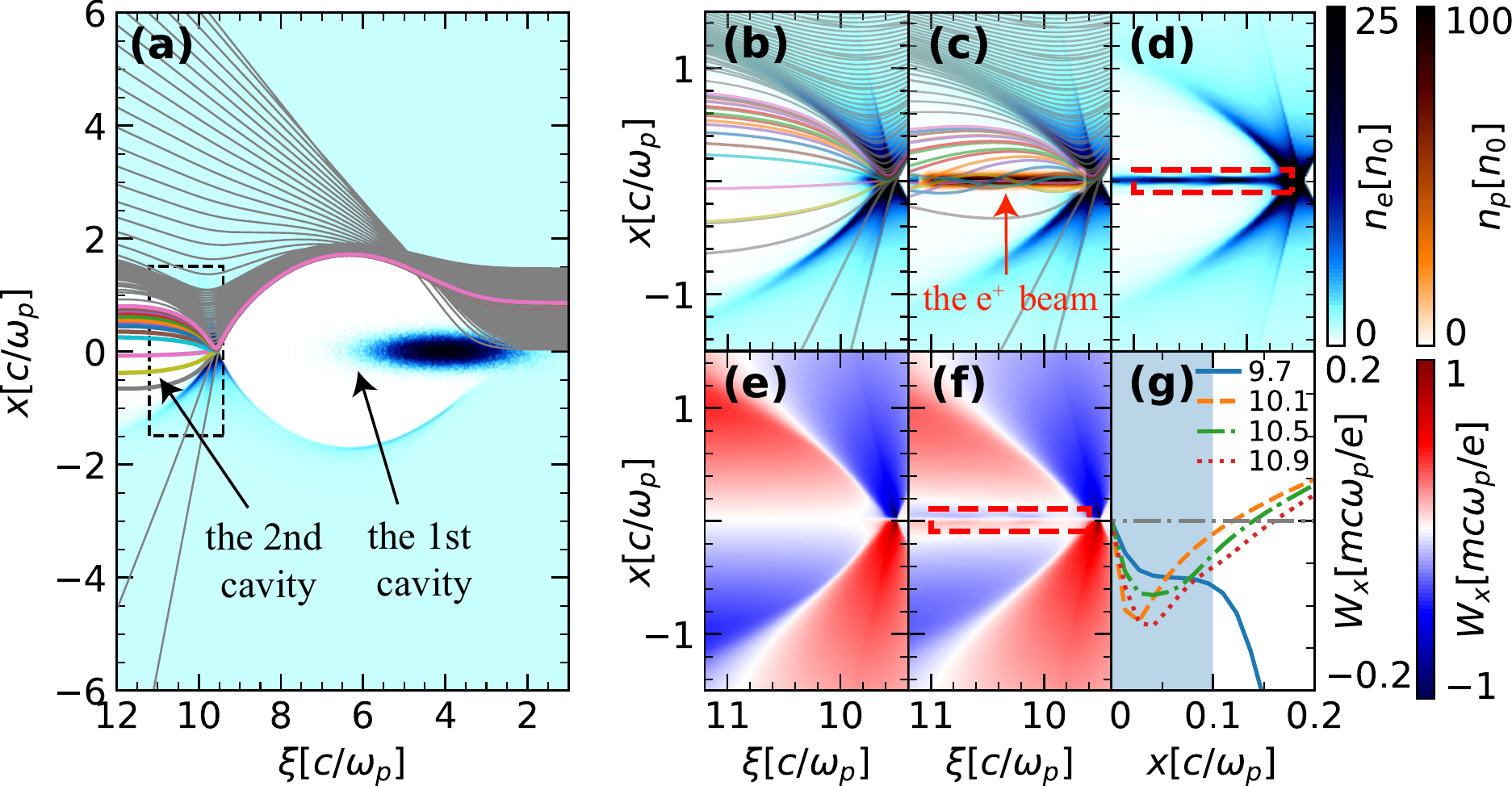}
  \caption{Beam density and trajectories of plasma electrons for (a) an unloaded blowout regime (see text below for the beam, plasma and simulation parameters), (b) the zoom-in-view of the dashed box and (c, d) positron beam loading case. Transverse wakefield $W_x\equiv E_x-cB_y$ for the (e) unloaded and (f) beam-loading case and (g) its transverse lineouts at $k_p\xi=9.7,10.1,10.5,10.9$. Red boxes indicate the position of positron beam.}
  \label{Fig:1}
\end{figure}

In the blowout regime of PWFA where the ions are uniformly distributed within the cavity, the transverse force for a relativistic positron ($v_z\sim c$) is\cite{Lu2006POP,Lu2006PRL}
\begin{gather}
  \begin{aligned}
    F_{\perp e^+}(r) &= e(E_r-v_z B_\theta)=-e\frac{\partial\psi}{\partial r}\\
    &= \frac{1}{2}r + \frac{1}{r}\int_{0}^{r} \left[\rho_e \left(r^\prime\right)- J_{ze}\left(r^\prime\right)/c \right]r^\prime dr^\prime,
  \end{aligned}
  \label{eq:1}
\end{gather}
where $\psi\equiv\phi-A_z$  is the wake pseudo-potential, $\phi$ and $\boldsymbol{A}$ are the scalar and vector electromagnetic potential, $\rho_e$ and $\boldsymbol{J}_e$ are the charge and current density of the plasma electrons and axial symmetry and quasi-static approximation\cite{Mora1997} are assumed. Henceforth, we adopt normalized units with length, speed, density, mass and charge normalized to the plasma skin-depth, $c/\omega_p$, speed of light, \textit{c}, plasma density, $n_0$, electron rest mass, \textit{m}, and electron charge, \textit{e}, respectively. In Eq.\eqref{eq:1}, the 1st term is the repulsive force from ions and the 2nd is focusing induced by the plasma electrons remaining within the cavity. To focus a positron beam with transverse size $r_0$,
\begin{equation}
  -\int_{0}^{r_0}{\left[\rho_e \left(r^\prime\right)-J_{ze} \left(r^\prime\right)/c\right]r^\prime}dr^\prime>\frac{1}{2}r_0^2
  \label{eq:2}
\end{equation}
must be satisfied. In an unloaded blowout wake, only the narrow volume between the first and second bubbles fulfills the criterion as in Fig.\ref{Fig:1}. However, placing a positron beam into this region dramatically affects the distribution of the plasma electrons around. The transverse force for a plasma electron with $v_z$ is
\begin{equation}
  \begin{aligned}
    F_{\perp e^-}\left(r\right)& =-\frac{r}{2}-\frac{1-v_z}{r}\int_{0}^{r}{n_br^\prime}dr^\prime-\frac{1}{r}\bigg [\int_{0}^{r}{\rho_er^\prime}dr^\prime\\ 
    & -v_z\int_{0}^{r}{\frac{J_{ze}}{c}r^\prime}dr^\prime \bigg ] - \frac{1-v_z}{2}\int_{0}^{r}\frac{\partial E_z}{\partial\xi}dr^\prime,
  \end{aligned}
  \label{eq:3}
\end{equation}
where $n_b$ is the witness positron beam density and $\xi\equiv ct-z$. The four terms that contribute to the transverse force are due to the ions, positron beam, $\rho_e$ and $J_{ze}$  and the plasma transverse current respectively. 

The plasma electron densities and transverse force distributions in a typical blowout regime with/without positron beam loading are visualized in Fig.\ref{Fig:1} based on PIC simulations and particle trackings. Figure \ref{Fig:1}a depicts the unloaded situation in ($x, \ \xi$) plane where $n_{driver}\gg n_0$, and $k_p\sigma_{r,z}\le1$. The plasma electrons are repelled by the relativistic drive electrons and then pulled back by the exposed ions, forming a bubble void of electrons. At the end of the bubble, most electrons are reflected due to the 3rd and 4th terms in Eq.\eqref{eq:3}\cite{Xu2017}. As these electrons are deflected, the focusing force provided by the immobile plasma ions (and the positron beam if placed as in Fig. \ref{Fig:1}c) becomes dominant, which results in some plasma electrons to bend and stay close to the axis in the second wake cavity to provide focusing force for positrons. If the positron beam plus the ions balance the self-repulsion force of sufficient sheath electrons, the latter form a coaxial filament and the positron beam can be confined.

Simulations are conducted using the quasi-static 3D-PIC code \textit{QuickPIC}\cite{Huang2006,An2013}, with a simulation window $14\times 14\times 15 k_p^{-3}(x,\ y,\ \xi)$ and resolution of $0.014\times 0.014\times 0.015 k_p^{-3}$. The drive beam has a bi-Gaussian profile centered at $k_p\xi=4$ with $k_p\sigma_z=1$, $k_p\sigma_r=0.17$, and $n_{driver}{/n}_0=23$. If the plasma density is $7.8\times{10}^{15}{\rm cm}^{-3}$ then $k_p^{-1}=60\mathrm{\mu m}$, corresponding to the drive beam containing 2.75nC charge with $\sigma_z=60\mathrm{\mu m}$, $\sigma_r=10\mathrm{\mu m}$. This driver excites a nonlinear blowout wake where the returning plasma electrons converge around the axis in a small volume. Figure \ref{Fig:1}e shows that in an unloaded case the region of the electron convergence (density peak) is the only focusing area for an on-axis positron beam. However, if a narrow positron beam (in the simulation it has a longitudinally flattop profile at $k_p\xi\in\left[9.6,\;11\right]$ and a transverse Gaussian profile of $\sigma_p=2\mathrm{\mu m}$ with a peak density 100$n_0$) is loaded into the cavity just behind the density peak as in Fig.\ref{Fig:1}c, the situation changes dramatically. We track the trajectories of selected plasma electrons whose original radii uniformly distribute between 0 to $1.5k_p^{-1}$. Electrons returning to the axis quickly diverge again in the unloaded situation (Fig.\ref{Fig:1}b), whereas the space charge of the positron beam attracts many of these electrons (Fig.\ref{Fig:1}c) to the axis, thereby creating an extended electron filament that overlaps with the positron bunch (Fig.\ref{Fig:1}d). Figure \ref{Fig:1}f shows that now there is a net focusing force on the positron beam. The transverse lineouts in Fig.\ref{Fig:1}g validate that the entire positron beam ($\pm3\sigma_p$) is focused, as expected from Eq.\eqref{eq:2}. Unfortunately, since the electron density in the filament is nonuniform, the transverse focusing field experienced by the positrons is nonlinear along \textit{r} and varies along $\xi$\cite{Zhou2021}. Emittance preservation in this situation can be achieved by approximate matching and slice-by-slice matching techniques\cite{Benedetti2017,Diederichs2020}. In general, the self-consistent focusing force requires narrow size and high intensity of the positron bunch. Otherwise the positron beam may suffer from some slices experiences a defocusing force and dispersion during propagation.

\begin{figure}[ht]
  \includegraphics[width=8.6cm]{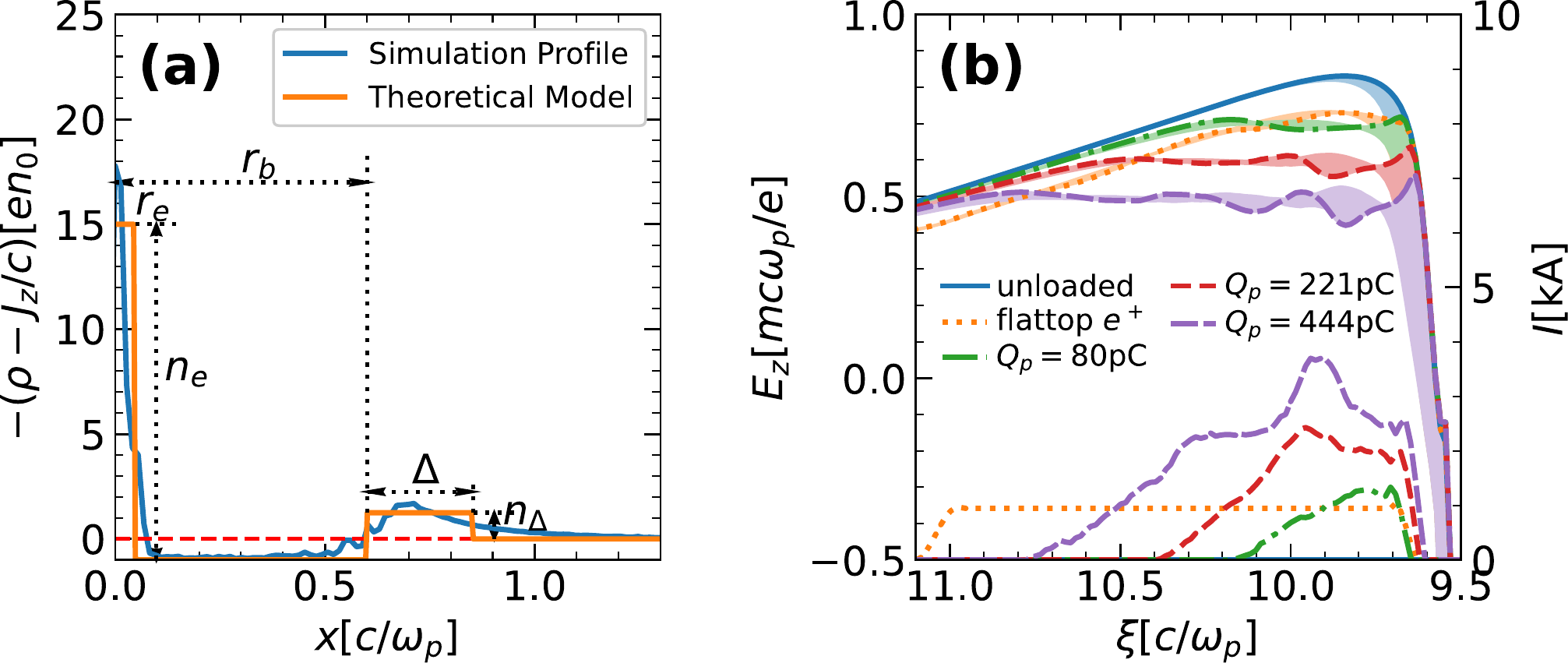}
  \caption{Beam loading effects in the coaxial bubble. (a) A transverse lineout of \textit{S} at $k_p\xi=10.3$ in simulation and theoretical model. (b) On-axis $E_z$ (line) and range of $E_z$ within $\pm3\sigma_p$ (shadow) for different optimized positron currents.}
  \label{Fig:2}
\end{figure}

Due to the coaxial electron filament inside the second cavity, the positron beam loading on the accelerating field differs from electron beam loading. Upon applying the quasi-static approximation, wakefields in PWFA are determined by the pseudo-potential $\psi$, which obeys two-dimensional Poisson equation
\begin{equation}
  \nabla_\bot^2 \psi = -\left(\rho-J_z/c\right)\equiv S,
  \label{eq:4}
\end{equation}
where$\nabla_\bot\equiv\hat{x}\frac{\partial}{\partial x}+\hat{y}\frac{\partial}{\partial y}$, $\rho-J_z/c=\rho_{ion}+\rho_e-J_{ze}/c-J_{zion}/c$ and we have $\rho_{ion}=1$, $J_{zion}=0$. Figure \ref{Fig:2}a illustrates a lineout of $S$ at $k_p\xi=10.3$ for the positron beam loading case of Fig.\ref{Fig:1}. In the second cavity, \textit{S} can be split into the electron filament, blowout region, electron sheath and the unperturbed plasma respectively from inside to outside similar to the blowout regime\cite{Lu2006PRL,Dalichaouch2021}. We approximate each part by step functions as presented in Fig.\ref{Fig:2}a.

The width and amplitude of the electron filament and sheath are $r_e$, $n_e$, $\mathrm{\Delta}$, $n_\mathrm{\Delta}$ respectively, and the blowout radius is $r_b$. Within each slice, the total \textit{S} is conserved leading to
\begin{equation}
  (n_e-1)r_e^2-\left(r_b^2-r_e^2\right)+n_\mathrm{\Delta}\left[\left(r_b+\mathrm{\Delta}\right)^2-r_b^2\right]=0
  \label{eq:5}
\end{equation}
With the piecewise-step approximation and Eq.\eqref{eq:5}, the on-axis pseudo-potential, $\psi_0\equiv\psi\left(r=0\right)$, can be obtained by solving Eq.\eqref{eq:4},
\begin{gather}
  \begin{aligned}
    \psi_0\left(\xi\right) &=\frac{1}{4}r_b^2\left(\xi\right)\left(1+\beta\left(\xi\right)\right)-\frac{1}{4}n_e\left(\xi\right)r_e^2\left(\xi\right)\left(1+\beta\left(\xi\right)\right)\\
    &-\frac{1}{2}n_e\left(\xi\right)r_e^2\left(\xi\right)\ln{\frac{r_b\left(\xi\right)}{r_e\left(\xi\right)}},
  \end{aligned}
  \label{eq:6}
\end{gather}
where $\beta\left(\xi\right)\equiv\frac{\left(1+\alpha\right)^2 \ln\left(1+\alpha\right)^2}{\left(1+\alpha\right)^2-1}-1$ and $\alpha\equiv\Delta / r_b$. Without the electron filament ($n_e=0$ or $r_e=0$) we recover $\psi_0=\frac{1}{4}r_b^2\left(1+\beta\right)$ for a normal blowout regime\cite{Lu2006PRL,Lu2006POP}. Assuming a uniform electron filament along $\xi$ ($\frac{\partial r_e}{\partial\xi}=\frac{\partial n_e}{\partial\xi}=0$) and $\frac{\partial\alpha}{\partial\xi}=0$, the on-axis $E_z$ can then be written as two terms,
\begin{gather}
  \begin{aligned}
    &E_{z0}\left(\xi\right)=\frac{{\partial\psi}_0\left(\xi\right)}{\partial\xi}\equiv E_{z1}+ E_{z2}\\
    =&\frac{\partial}{\partial\xi}\left[\frac{r_b^2\left(\xi\right)}{4}\left(1+\beta\right)\right]+ \left[-\frac{1}{2}n_er_e^2\frac{r_b'\left(\xi\right)}{r_b\left(\xi\right)} \right].
  \end{aligned}
  \label{eq:7}
\end{gather}
In Eq.\eqref{eq:7}, $E_{z1}$ is determined by the shape of the loaded bubble cavity just like in the case of electron beam loading\cite{Tzoufras2008}, and $E_{z2}$ is a novel effect related to both the electron filament and the bubble profile. The first term is roughly proportional to $\frac{1}{2}r_br_b^\prime$ which decreases with $\xi$ at the front of the second bubble in the absence of witness beam, as seen in Fig.\ref{Fig:2}b. The space charge of the positron beam decreases both $r_b$ and $r_b^\prime$ and thus more rapidly decreases $E_{z1}$ as $\xi$ increases, making it impossible to flatten the wake in this region. However, the electron filament modifies the load in two ways. First, its space charge will cancel that of the positron beam thereby decrease the effect on $r_b$. Second, $E_{z2}$ is of the opposite sign with $E_{z1}$ and it also decreases in magnitude as $\xi$ increases since $1/r_b$ and $r_b'$ damp as the bubble expands, which is exactly what is needed to flatten the wake. The trend to flatten the wake as the current of the positron beam increases is shown in Fig. \ref{Fig:2}b.

However, electrons in the filament execute betatron oscillations due to the strong focusing force provided by the $e^+$ beam, which causes longitudinal variations in the filament. According to the Panofsky-Wenzel theorem,
\begin{gather}
  \begin{aligned}
    &\frac{\partial E_z\left(r,\xi\right)}{\partial r}=\frac{\partial W_\bot\left(r,\xi\right)}{\partial\xi}=\frac{\partial\psi\left(r,\xi\right)}{\partial r\partial\xi}\\
    =&\frac{\partial}{\partial\xi}\left\lbrace -\frac{1}{r}\int_{0}^{r}{\left[\rho_e\left(r^\prime,\xi\right)-J_{ze}\left(r^\prime,\xi\right)/c\right]r^\prime}dr^\prime\right\rbrace,
  \end{aligned}
  \label{eq:8}
\end{gather}
longitudinal variations of the density filament (focusing force) lead to the transverse variation of $E_z$. In order for a transversely uniform $E_z$, $\rho_e-J_{ze}/c$ must not vary along $\xi$. To illustrate the transverse variation, we display the range of $E_z$ within $\pm3\sigma_{p}$ in Fig.\ref{Fig:2}b by the shaded regions. For the $e^+$ beam with flattop current profile, the transverse variation is small compared to the on-axis value except near the beam head.

Thus, a uniform average accelerating field $\left\langle E_z\right\rangle=\frac{\int_{0}^{\infty}E_z\left(r\right)n_b\left(r\right)rdr}{\int_{0}^{\infty}n_b\left(r\right)rdr}$ along the $e^+$ beam is essential for $e^+$ beam quality optimization. The optimized beam current profile can be obtained through an iterative algorithm similar to that described in \cite{Diederichs2020} assuming non-evolving drive and witness beams. Here the $e^+$ beam has a Gaussian transverse profile of $\sigma_p=2\mathrm{\mu m}$. The required current profiles and corresponding $E_z$ are presented in Fig.\ref{Fig:2}b. To extract more energy from the plasma wake, a higher charge is needed and the loaded $E_z$ is lower. However, owing to the nonlinear response of plasma electrons, the optimized $e^+$ beam current profile shapes change and the transverse variation of $E_z$ increases with the loaded charge because of the more intense effects described by Eq.\eqref{eq:8}. Therefore, for the 80, 221, 444pC $e^+$ beams, $\left\langle E_z\right\rangle$ is 0.7, 0.6, 0.5$mc\omega_p/e$ respectively, while the induced rms energy spread $\sigma_\gamma/\left(\gamma-\gamma_0\right)$ is 1.66\%, 3.30\% and 4.56\%, which indicates a tradeoff between the beam charge, efficiency, gradient and beam quality for positron acceleration in the blowout regime\cite{Hue2021}.

\begin{figure}[ht]
  \includegraphics[width=8.6cm]{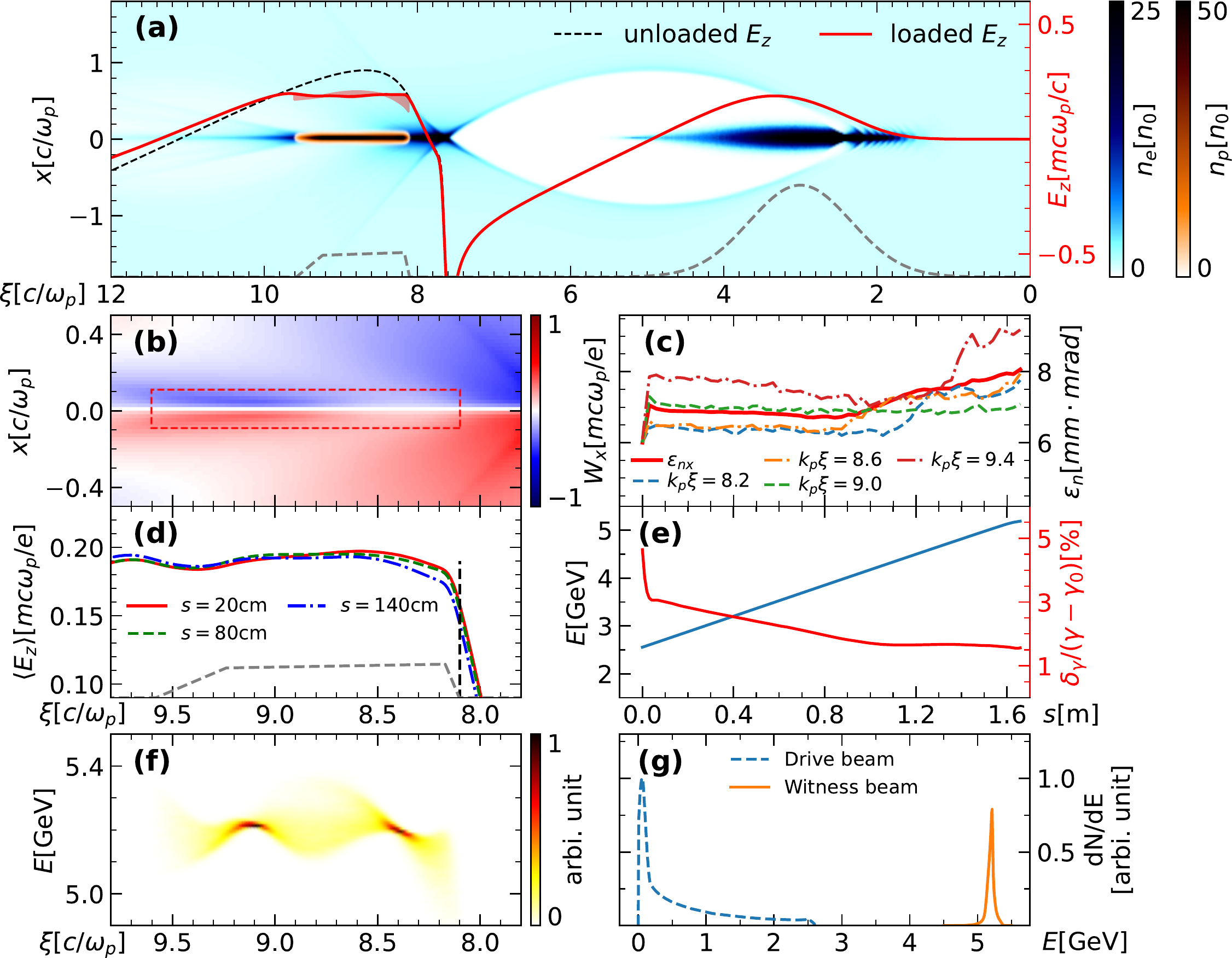}
  \caption{Results of positron beam acceleration in the blowout regime. (a) A snapshot of the plasma wake after propagating 20cm. The red shadow denotes $E_z$ within $\pm3\sigma_p$. (b) The distribution of $W_x$ where the rectangle presents the positron position. (c) Evolution of the slice and projected normalized emittances. (d) Average accelerating field $\left\langle E_z\right\rangle$ for the positron beam at different distances. (e) Evolution of the mean energy and induced energy spread. (f) Final longitudinal phasespace. (g) Final spectra of the two beams. The gray dashed lines show the beam current profile.}
  \label{Fig:3}
\end{figure}

It is still possible to achieve high efficiency $e^+$ beam acceleration with good beam quality over a long propagation distance. For example, in Fig.\ref{Fig:3} we present a case with $n_0 = 7.8\times{10}^{15}\mathrm{cm}^{-3}$, a bi-Gaussian electron driver of $\sigma_z=40\mathrm{\mu m}$, $\sigma_{r}=5\mathrm{\mu m}$ and normalized emittance $2.5 \mathrm{mm\cdot mrad}$. This beam contains 534pC charge and drives a bubble with a maximum radius of $\sim k_p^{-1}$. A positron beam with 102pC of charge, transverse size $\sigma_{p}=2\mathrm{\mu m}$, $\epsilon_n = 6\mathrm{mm\cdot mrad}$ and a piecewise-linear current profile is loaded at a distance $306\mathrm{\mu m}$ behind the drive beam center. Its current rises to $I_0=425\mathrm{A}$ in $4.8\mathrm{\mu m}$, decreases to 0.89$I_0$ in $64\mathrm{\mu m}$, then falls to 0 in 22$\mathrm{\mu m}$. Both beams have initial energy 2.5GeV ($\gamma_0=5000$). 

Figure \ref{Fig:3}a illustrates a snapshot of the plasma, beam densities and the corresponding $E_z$ at a propagation distance of 20cm. The $e^+$ beam loaded $E_z$ is almost flattened, and its transverse variation within $\pm 3\sigma_{p}$ is also suppressed. The corresponding transverse wakefield presented in Fig.\ref{Fig:3}b shows that the entire $e^+$ beam (indicated by the dashed rectangle) locates in a region of focusing fields which varies along $\xi$. As a result, beam emittances at different slices evolve differently as plotted in Fig.\ref{Fig:3}c, which rapidly grow at the beginning then saturate at various levels. After propagation of around 80cm, the emittances slowly increase again mainly because of the evolution of the driver. Finally, the projected emittance for the positron beam grows to about $8\mathrm{mm\cdot mrad}$ at $s= 165\mathrm{cm}$ and the results are almost the same for \textit{y} direction. The evolution or head erosion of the driver changes the focusing force for $e^+$ beam and causes emittance growth that cannot be mitigated by the above-mentioned matching techniques\cite{Benedetti2017,Diederichs2020} and can be suppressed by using a drive beam with emittance much less than the matched case as in our example.

The evolution of $\left\langle E_z\right\rangle$ for the $e^+$ beam is presented in Fig.\ref{Fig:3}d, which changes in front of the $e^+$ beam ($k_p\xi<8.5$) then varies little over the rest. Accordingly, the mean energy of the $e^+$ beam increases linearly, while the induced energy spread $\mathrm{\Delta}\equiv\delta_\gamma/\left(\gamma-\gamma_0\right)$ gradually evolves during acceleration as in Fig.\ref{Fig:3}e. Here, $\delta_\gamma\equiv1.48\delta_\gamma^{mad}$ and $\delta_\gamma^{mad}$ is the median absolute deviation of beam energy. This definition of the spread coincides with the rms value in case of a Gaussian distribution and is more robust for distributions deviating from Gaussian. After 165cm propagation, the $e^+$ beam is accelerated to 5.19GeV, corresponding to a mean gradient of 1.6GeV/m and $\mathrm{\Delta}$=1.56\% (induced rms energy spread $\mathrm{\Delta}_{rms}$=2.39\%). The final longitudinal phasespace for the $e^+$ beam (Fig.\ref{Fig:3}f) is consistent with the structure of $E_z$. The final spectra (Fig.\ref{Fig:3}g) show nearly complete energy depletion of some drive electrons and energy transfer efficiency from the wake to the positron beam $\eta=Q_pE_{p}^+/Q_eE_{e}^-$ is 26\%.

\begin{figure}[ht]
  \includegraphics[width=8.6cm]{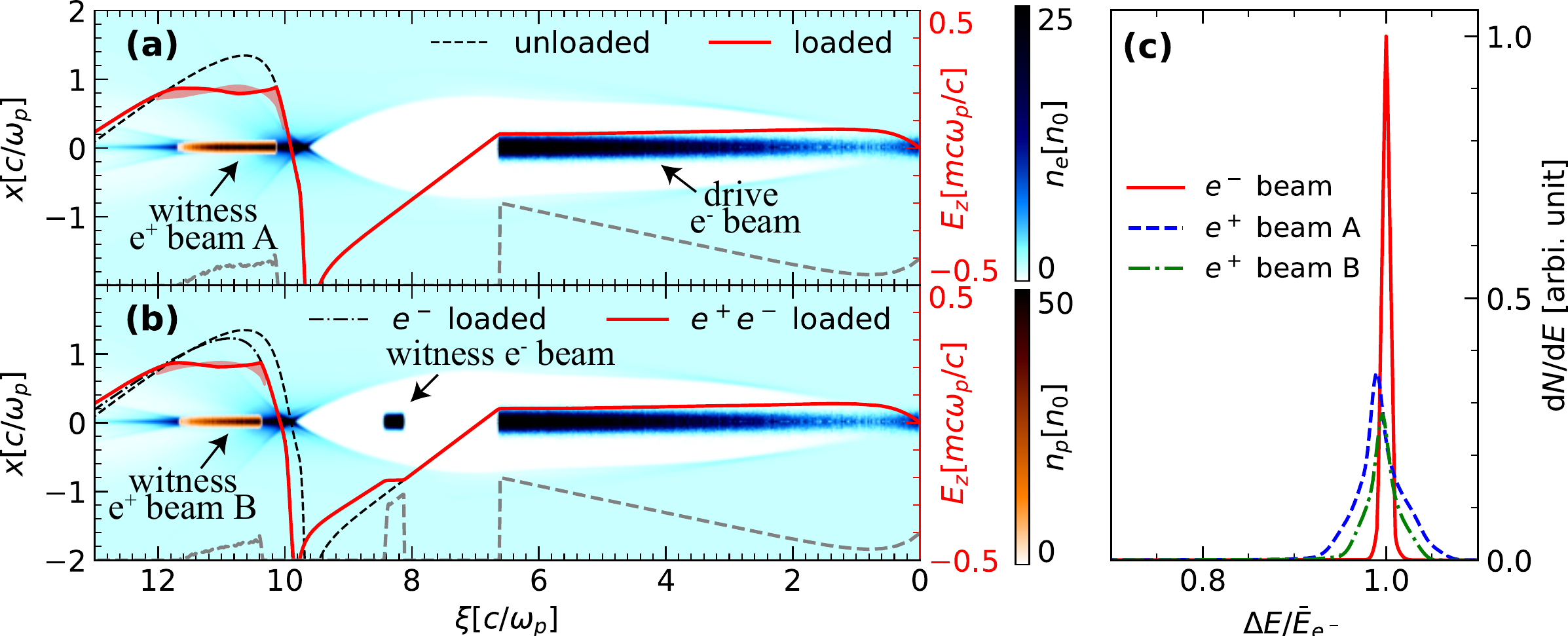}
  \caption{Configurations of the HTR (a) positron and (b) electron-positron acceleration. Gray dashed lines denote the beam current profile. (c) Spectra of the accelerated witness beams.}
  \label{Fig:4}
\end{figure}

Lastly, $e^+$ beam acceleration in the blowout regime opens new possibilities for PWFA. In Fig.\ref{Fig:4}a, we present an HTR positron acceleration case in a customized $e^-$ beam driven blowout regime\cite{Bane1985,Lu2009}. The plasma density is the same as above, the driver has transverse size $5\mathrm{\mu m}$ and current profile $I\left(\xi\right)\propto2k_p\xi-1+5e^{-2k_p\xi}$ with total length $400\mathrm{\mu m}$ and charge 1.4nC. Assuming non-evolving beams, an optimized $e^+$ beam obtained through the procedure in \cite{Diederichs2020} is presented. The positron beam has transverse size $2\mathrm{\mu m}$, contains 131pC charge, and experiences an average accelerating field of 1.78GV/m that is 3 times the maximum decelerating field of the driver. The induced rms energy spread is 2.75\% and the beam load efficiency is 35.3\%. Furthermore, an electron beam can also be loaded in the first bucket of the same wake to achieve simultaneously HTR $e^+e^-$ acceleration as in Fig.\ref{Fig:4}b. The witness $e^-$ and $e^+$ beams contain similar charge with the same accelerating gradient. By loading an electron beam, the energy spread for $e^+$ beam is improved due to the less intense loaded wake while the overall energy efficiency 44.1\% is increased compared to 35.3\% as listed in Table \ref{Table:1}. The transverse parameters of the $e^-$ beam are free variables and the matched $e^+$ beam emittance is several $\mathrm{mm\cdot mrad}$. The spectra for the witness beams are plotted in Fig.\ref{Fig:4}c. Compared with the $e^-$ beam, $e^+$ beams obtain a larger energy spread and the distributions deviate from Gaussian.

\begin{table}[ht]
  \caption{Parameters of the witness beams in Fig.\ref{Fig:4}.}
  \begin{tabular}{c|cccc}
    \hline
    witness beam & $Q$[pC] & $q\overline{E}$[GeV/m] & $\Delta_{rms}$[\%] & $\eta$[\%]\\
    \hline
    $e^+$ beam A & 131 & 1.78 & 2.75 & 35.3\\
    $e^+$ beam B & 80.4 & 1.78 & 2.03 & 21.3\\
    $e^-$ beam & 84.0 & 1.78 & 0.29 & 22.8\\
    \hline
  \end{tabular}
  \label{Table:1}
\end{table}

In summary, we have analyzed the positron beam loading in the blowout regime via PIC simulations and a phenomenological theoretical model and demonstrated its capability for high-quality positron acceleration with tens of percent energy transfer efficiency. The physics of positron beam loading is strongly determined by the dynamics of plasma electrons that can form an electron filament that overlaps with the positrons, and the beam quality is sensitive to the distribution of electrons in the filament. We have further shown that it is possible to simultaneously beam load an electron and a positron beam in a nonlinear blowout wake driven by an electron beam, and to obtain high gradient and efficiency acceleration of both beams. These ideas may be applied to laser driven nonlinear wakes as well. Further refinement of the theoretical model is possible which will lead to improvement of positron beam acceleration in plasma accelerators.

\bibliographystyle{apsrev4-2}
\bibliography{apssamp}

\end{document}